# Perceptually Relevant Preservation of Interaural Time Differences in Binaural Hearing Aids

Fábio P. Itturriet, and Márcio H. Costa, *Member, IEEE*

*Abstract*—This work presents a noise reduction method with perceptually relevant preservation of the interaural time difference (ITD) of the residual noise in binaural hearing aids. The interaural coherence (IC) concept, previously applied to the Multichannel Wiener Filter (MWF) for preservation of the spatial subjective sensation of diffuse noise fields, is proposed here to both preserve and emphasize the ITD binaural cues of a directional acoustic noise source. It is demonstrated that the previously developed MWF-ITD technique may decrease the original IC magnitude of the processed noise, consequently increasing the variance of the interaural phase difference (IPD) of the output signals. It is shown that the MWF-IC technique concomitantly minimizes a nonlinear function of the difference between input and output IPD, which is strictly related to ITD, and preserves the natural coherence of the directional noise captured by the reference microphones. Objective measures and psychoacoustic experiments corroborate the theoretical findings, showing the MWF-IC technique provides relevant noise reduction, while preserving the original ITD subjective perception and original lateralization for a directional noise source. These results are especially relevant for hearing aid designers, since they indicate the MWF-IC as a noise reduction technique that provides residual noise spatial preservation for both diffuse and directional noise sources in frequencies below 1.5 kHz.

## I. INTRODUCTION

NOISE reduction algorithms are important part of modern hearing aids. One of the major complaints of hearing aid users is poor speech intelligibility due to background noise. Many studies have demonstrated that hearing-impaired people need a SNR-50[1] from 10 to 30 dB higher than that required for the non-impaired [1]. This happens due to the loss of spectral resolution of the damaged auditory system [2]. Consequently, it may result in social isolation, professional difficulties and risk to personal safety. According to [3], more than 80 percent of the hearing impaired have both ears affected by a reduction in hearing ability, requiring the concomitant use of two hearing aids.

Bilateral hearing aids (left and right gadgets working independently) do not preserve the original acoustic localization cues, distorting the listener's sense of auditory space, as well as its ability to localize, separate, and track sound sources [4].

Although noise-reduction could be effective, localization of residual sounds is generally best achieved by turning off the processing routines [2] [4], diminishing the equipment acceptability. This represents a major disadvantage to the hearing-aid user since the immediate localization of sources of interest is paramount to allow visual identification (traffic, safety warnings) and/or lip-reading.

Despite many advances in hearing assistive technology, noise reduction strategies that preserve spatial localization of sound sources are still a challenging task, mainly due to the difficulty of integrating the different localization cues into the noise reduction framework. In this context, common approaches for noise reduction are the linearly constrained minimum variance beamformer and the generalized side-lobe canceller. However, these techniques rely on prior knowledge about source localization and/or head related transfer functions, presenting significant performance degradation when the assumed conditions deviate from the real ones [5].

Binaural[2] Multichannel Wiener Filter (MWF) based techniques have been extensively explored in the current scientific literature [6] [7] [8] [9] [10]. This approach permits deep theoretical insights about its design and performance [11] [12] [13]. Although it was theoretically demonstrated that the conventional binaural MWF naturally preserves speech localization cues, its major drawback resides in the fact that residual noise at the output inherits the input speech localization cues [11]. As a result, the hearing aid user cannot make use of the psychoacoustic mechanisms related to spatial separation between noise and speech sources [14] to mask unwanted information (best ear advantage) and therefore improve the speech understanding or to localize/track noise sources [6] [15].

In order to overcome such issue, some MWF variations have been developed. They can be divided in two classes: In the first one, controlled amounts of unprocessed signal are allowed at the output of the hearing aids. Although it was demonstrated that it results a better noise source spatial localization as compared to the conventional MWF [15] [16], this paradigm is not strictly related to preservation of the localization cues. The second approach is characterized by adding extra terms to the MWF cost function to penalize solutions that do not preserve the desired binaural cues. It has been demonstrated that interaural time differences (ITD) (difference between transmission times in both ears) are the primary spatial cues in mammals and birds [17], followed by interaural level differences (ILD) (difference between magnitudes at each ear). Other cues are also accepted as playing secondary role (providing supplementary information) in sound source localization, such as monaural spectral features provided by pinnae, and non-acoustical cues such as source familiarity and vision [18]. For frequency components above 1500 Hz, ITD may exhibit ambiguity due to short wavelengths compared to the distance between the ears, and due to the breakdown of phase locking in the auditory neurons [19] [20] [21]. ILD is mainly determined by the attenuation provided by the head and torso (head shadow effect) and is more pronounced when the head is in between the source and one of the ears. Due to the acoustic properties (reflection, diffraction, scattering, interference and resonance) of the head, torso and pinnae, ILD presents a strongly increasing dependence with frequency [19] [22]. When available, low-frequency ITD information is dominant over ILD and spectral shape information, which are used to resolve ambiguities [18].

---

[1] SNR-50 is the signal to noise ratio needed for the comprehension of 50% of the speech in a conversation.

[2] When both (left and right) hearing aids exchange data.

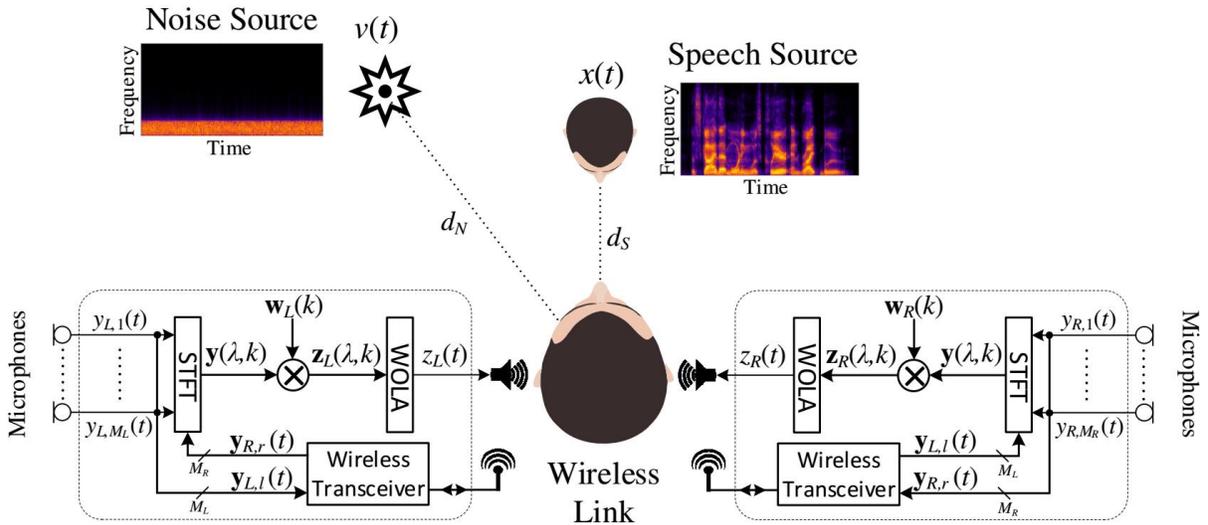

Fig. 1. Binaural system setup.

The first attempt to preserve ITD and ILD of residual noise by means of inserting an auxiliary cost function into the MWF formulation was presented in [7]. The efficacy of MWF-ILD methods [7] [23] for preservation of the acoustic scenario was supported by early amplitude stereo panning techniques that have demonstrated that ILD carries enough information for creating complex artificial auditory scenes even in headphones [24]. Differently from the MWF-ILD, and despite the apparent physical appropriateness of the proposed estimator and associated cost function, extensive experiments have demonstrated that the MWF-ITD [25] does not provide perceptually relevant results in preserving the localization of the residual noise. This fact can be only partially explained by the observations presented in [26]. As a result, this is still an important open research area.

In this work, we propose the use of the interaural coherence (IC) measure for preserving and emphasizing the ITD localization cues in MWF processed signals. The IC was firstly proposed in [27] for preserving the original spatial characteristics of diffuse noise fields, and in [28] it was shown the IC is a nonlinear function of the ITD. Here, it is shown that by minimizing IC differences between input and processed signals the original ITD localization cues are also preserved, resulting in the correct psychoacoustic impression of the original acoustic scene for directional sources. The novel contributions of this work are: 1) It is provided strong experimental evidence and theoretical support that the MWF-ITD technique, as originally described in [7], is not capable of preserving the original lateralization of the processed noise; 2) It is shown that minimizing the difference between input and output IC of a signal produced by a directional acoustic source corresponds to minimize the difference between input and output ITD; 3) We propose the MWF-IC technique, originally derived for diffuse fields, as an efficient noise reduction method for providing ITD preservation of directional sound sources; 4) It is provided objective results and psychoacoustic experiments that corroborate the previous rationales, showing that the MWF-IC noise reduction technique leads to perceptually relevant preservation of the ITD localization cues.

The remainder of this paper is structured as follows: the binaural problem formulation is presented in Section II, while Section III introduces the MWF noise reduction technique. A brief review of binaural cost functions for preserving the binaural cues is presented in Section IV. In Section V, the MWF-IC technique is proposed as an efficient noise reduction technique with ITD preservation of directional sources. The experimental setup is described in Section VI, and the results are presented in Section VII. Finally, Section VIII and IX present the discussion and conclusions of this work.

Throughout this document, bold uppercase and lowercase letters represent matrices and vectors, respectively, while italics represent scalars.

## II. SIGNAL AND SYSTEM MODELS

The application context considered in this paper comprises a binaural fitting of hearing aids, working in a full-duplex mode without bit-rate limitations. The operating scenario assumes the existence of one acoustic source of interest $x(t)$ (speech) and one interfering noise source $v(t)$. Both sources are assumed as having a fixed (or slowly varying) position in a given time-window. Frequency domain decomposition is applied to the incoming signals through an *N*-bin Short-Time Fourier Transform (STFT). For a sampling frequency of $f_s$ samples per second, for each time-frame $\lambda$ and frequency $k$, for the $M_L$ left microphones and the $M_R$ right microphones, the received signals are defined as:

$$y_{L,l}(\lambda,k) = x_{L,l}(\lambda,k) + v_{L,l}(\lambda,k) \\ y_{R,r}(\lambda,k) = x_{R,r}(\lambda,k) + v_{R,r}(\lambda,k)$$ , (1)

in which $x$ is the speech signal, $v$ is the noise, $l = 1,\ldots,M_L$, and $r = 1,\ldots,M_R$. The collection of these received signals can be expressed in vector form as

$$\mathbf{y}(\lambda,k) = \mathbf{x}(\lambda,k) + \mathbf{v}(\lambda,k) ,$$ (2)

in which $\mathbf{y}(\lambda,k) = [\ y_{L,1}(\lambda,k) \ \ldots \ y_{L,M_L}(\lambda,k) \ y_{R,1}(\lambda,k)\ldots \ y_{R,M_R}(\lambda,k)\ ]^T$, $\mathbf{x}(\lambda,k) = [\ x_{L,1}(\lambda,k) \ \ldots \ x_{L,M_L}(\lambda,k) \ x_{R,1}(\lambda,k)\ldots \ x_{R,M_R}(\lambda,k)\ ]^T$ and $\mathbf{v}(\lambda,k) = [\ v_{L,1}(\lambda,k) \ \ldots \ v_{L,M_L}(\lambda,k) \ v_{R,1}(\lambda,k)\ldots \ v_{R,M_R}(\lambda,k)\ ]^T$ are vectors with dimension $M \times 1$ and $M = M_L + M_R$.

Considering the deterministic vectors $\mathbf{q}_L$ and $\mathbf{q}_R$, both with dimensions $M \times 1$, which contain 1 in the element corresponding to the respective (left/right) reference microphone and zeros in all other elements, the reference vectors of the hearing aids (without processing) are given by

$$y_{L,ref}(\lambda,k) = x_{L,ref}(\lambda,k) + v_{L,ref}(\lambda,k) = \mathbf{q}_L^T \mathbf{y}(\lambda,k) \\ y_{R,ref}(\lambda,k) = x_{R,ref}(\lambda,k) + v_{R,ref}(\lambda,k) = \mathbf{q}_R^T \mathbf{y}(\lambda,k)$$ . (3)

As shown in Fig. 1 the output signals of the hearing aids are

$$z_L(\lambda,k) = \mathbf{w}_L^H(\lambda,k)\mathbf{y}(\lambda,k) \\ z_R(\lambda,k) = \mathbf{w}_R^H(\lambda,k)\mathbf{y}(\lambda,k)$$ , (4)

where $\mathbf{w}_L(\lambda,k)$ and $\mathbf{w}_R(\lambda,k)$ are the left and right coefficient vectors of the noise reduction multichannel filter, both with dimension $M \times 1$.

## III. MULTICHANNEL WIENER FILTER

The binaural MWF has been largely studied in the noise reduction context for hearing aid applications. Its cost function is given by [29]

$$J_W(k) = \mathbb{E}\left\{\left\|\begin{matrix}x_{L,ref}(\lambda,k) - \mathbf{w}_L^H(k)\mathbf{y}(\lambda,k)\\ x_{R,ref}(\lambda,k) - \mathbf{w}_R^H(k)\mathbf{y}(\lambda,k)\end{matrix}\right\|^2\right\},$$ (5)

where $\mathbb{E}\{\cdot\}$ indicates the expected value and $\|\cdot\|^2$ is the squared Euclidean norm. Manipulating (5) leads to [30]

$$J_W(k) = \mathbf{q}_L^T\mathbf{\Phi}_{\mathbf{xx}}(k)\mathbf{q}_L + \mathbf{q}_R^T\mathbf{\Phi}_{\mathbf{xx}}(k)\mathbf{q}_R - \mathbf{q}_L^T\mathbf{\Phi}_{\mathbf{xx}}(k)\mathbf{w}_L(k) \\ -\mathbf{q}_R^T\mathbf{\Phi}_{\mathbf{xx}}(k)\mathbf{w}_R(k) - \mathbf{w}_L^H(k)\mathbf{\Phi}_{\mathbf{xx}}(k)\mathbf{q}_L \\ -\mathbf{w}_R^H(k)\mathbf{\Phi}_{\mathbf{xx}}(k)\mathbf{q}_R + \mathbf{w}_L^H(k)\mathbf{\Phi}_{\mathbf{yy}}(k)\mathbf{w}_L(k) \\ +\mathbf{w}_R^H(k)\mathbf{\Phi}_{\mathbf{yy}}(k)\mathbf{w}_R(k)$$ , (6)

in which coherence matrices $\mathbf{\Phi}_{\mathbf{xx}}(k) = \mathbb{E}\{\mathbf{x}(\lambda,k)\mathbf{x}^H(\lambda,k)\}$ and $\mathbf{\Phi}_{\mathbf{yy}}(k) = \mathbb{E}\{\mathbf{y}(\lambda,k)\mathbf{y}^H(\lambda,k)\}$ are assumed Hermitian positive semi-definite. Equation (6) is a quadratic function of the coefficient vectors $\mathbf{w}_L(k)$ and $\mathbf{w}_R(k)$. Due to its strict convexity, the minimum of $J_W(k)$ is found in closed-form by equating to zero its partial derivatives w.r.t. the coefficients. It was shown that the use of the obtained coefficient vectors in the system shown in Fig. 1 provides significant noise reduction and speech source spatial preservation [31].

## IV. BINAURAL COST FUNCTIONS

It was previously discussed that the binaural MWF distorts the perception of the noise source localization [11]. In order to provide a trade-off between noise reduction and spatial preservation, auxiliary cost functions have been proposed in the literature [6] [7] [8] [27]. These cost functions are combined with $J_W(k)$ and can be generalized by:

$$J(k) = J_W(k) + \sum_i \alpha_i(k) J_i^v(k) ,$$ (7)

in which $J(k)$ is the cost function to be minimized w.r.t. $\mathbf{w}_R(k)$ and $\mathbf{w}_L(k)$; $J_W(k)$ is the multichannel Wiener filter cost function, responsible for noise reduction; $J_i^v(k)$ are a set of auxiliary cost functions which aim to preserve the noise binaural cues; and $i \in \{\text{ITD, ILD, ITF}\}$ for a directional noise source or $i \in \{\text{IC}\}$ for diffuse noise. Parameters $\alpha_i(k)$ are frequency dependent weighting factors that take into consideration the importance of preservation of binaural cues as compared to the noise reduction effort. Each auxiliary cost function is defined as the difference between input and output estimates of a given binaural cue. Optimization techniques are applied to (7) for finding the optimum (left/right) coefficient vectors that minimize $J(k)$ for each bin.

### A. Interaural Time Difference

The ITD (in seconds) was defined in [32] as the phase of the ratio between the left and right signal components in the reference microphone. The input noise ITD at each bin and time-frame is defined as

$$ITD_{in}^v(\lambda,k) = \frac{1}{2\pi f_s k} IPD_{in}^v(\lambda,k) \\ = \frac{1}{2\pi f_s k}\left(\angle \frac{v_{L,ref}(\lambda,k)}{v_{R,ref}(\lambda,k)} + 2\pi p(k)\right)$$ , (8)

where $\angle$ means phase of its argument. The integer $p(k)$ is the phase unwrapping factor, which is a *priori* unknown, since the angle of the ratio of the spectra is computed modulo $2\pi$. This makes the phase ambiguous above 1500 Hz due to the size and shape of the human head. For frequencies below 1500 Hz, $p(k)$ can be considered zero [32]. Under such condition, the mean input noise IPD (in radians), at a given time-window, can be calculated by the following approximation

$$IPD_{in}^v(k) = \mathbb{E}\left\{\angle \frac{v_{L,ref}(\lambda,k)}{v_{R,ref}(\lambda,k)} \times \frac{v_{R,ref}^*(\lambda,k)}{v_{R,ref}^*(\lambda,k)}\right\} \\ = \mathbb{E}\left\{\angle v_{L,ref}(\lambda,k)v_{R,ref}^*(\lambda,k)\right\} \\ \cong \angle \mathbb{E}\left\{v_{L,ref}(\lambda,k)v_{R,ref}^*(\lambda,k)\right\} \\ = \angle \mathbf{q}_L^T \mathbf{\Phi}_{\mathbf{vv}}(k) \mathbf{q}_R$$ , (9)

where $\mathbf{\Phi}_{\mathbf{vv}}(k) = \mathbb{E}\{\mathbf{v}(\lambda,k)\mathbf{v}^H(\lambda,k)\}$. Using the same approach for $z_L(\lambda,k)$ and $z_R(\lambda,k)$ results the mean output noise IPD:

$$IPD_{out}^v(k) = \angle \mathbf{w}_L^T(k)\mathbf{\Phi}_{\mathbf{vv}}(k)\mathbf{w}_R(k),$$ (10)

which is defined as the phase difference between the output signals in both speakers. Using (9) and (10), the ITD cost function is defined as:

$$J_{ITD}^v(k) = |ITD_{out}^v(k) - ITD_{in}^v(k)|^2$$
$$= \frac{1}{4\pi^2 f_s^2 k^2} |IPD_{out}^v(k) - IPD_{in}^v(k)|^2, \quad (11)$$
$$= \frac{1}{4\pi^2 f_s^2 k^2} J_{IPD}^v(k)$$

resulting in the MWF-ITD cost function [7]:

$$J_T(k) = J_W(k) + \alpha_T(k) J_{IPD}^v(k), \quad (12)$$

in which the constant $(2\pi f_s k)^{-2}$ was included into $\alpha_T(k)$.

## V. PROPOSED METHOD

In this section, we firstly analyze the accuracy of the mean IPD as an estimator of the spatial azimuth for a directional acoustic source. Following, we propose the use of the IC for preserving the original spatialization of directional noise sources in MWF based noise reduction systems for hearing aids.

### A. Performance Analysis of the IPD Estimator

It is reasonable to assume by the central limit theorem [33] that, in the STFT domain, the input noise components at both microphones, $v_L(\lambda,k)$ and $v_R(\lambda,k)$, are zero-mean complex random variables, normally distributed, with zero mean and coherence matrix given by

$$\boldsymbol{\Phi}_{\mathbf{v}_L \mathbf{v}_R}(k) = \begin{bmatrix} \sigma_{v_L}^2(k) & \rho(k)\sigma_{v_L}(k)\sigma_{v_R}(k) \\ \rho^*(k)\sigma_{v_L}(k)\sigma_{v_R}(k) & \sigma_{v_R}^2(k) \end{bmatrix}, \quad (13)$$

in which $\sigma_{v_L}^2(k) = \mathbb{E}\{|v_L(\lambda,k)|^2\}$, $\sigma_{v_R}^2(k) = \mathbb{E}\{|v_R(\lambda,k)|^2\}$, and $\rho(k) = \mathbb{E}\{v_L(\lambda,k)v_R(\lambda,k)^*\}/[\sigma_{v_L}(k)\sigma_{v_R}(k)]$ is the complex coherence coefficient of the left and right reference microphones. The probability density function of the random variable $\theta = \angle v_L(\lambda,k)/v_R(\lambda,k)$ (which is equal to $IPD_{in}^v(\lambda,k)$ for $\omega < 1500$ Hz) for each bin and time-frame is given by:

$$p_\theta(\theta) = \frac{1-|\rho|^2}{2\pi(1-\eta^2)} \left[ \frac{\eta}{\sqrt{1-\eta^2}} \arccos(-\eta) + 1 \right], \quad (14)$$

in which $\eta = |\rho|\cos(\angle\rho - \theta)$ (see Appendix).

Fig. 2 plots Eq. (14) for $\rho = |\rho|\exp(j\pi/4)$ and different values of $|\rho|$. It can be observed that the central tendency ($\pi/4$) remains fixed, but the dispersion increases with the decrease of the absolute value of the coherence coefficient $|\rho|$. In low reverberation scenarios, directional noise captured by the reference microphones naturally presents large coherence coefficient (usually near one). Therefore, $IPD_{in}^v(k)$ results accurate estimates for $\mathbb{E}\{\theta\}$. By the other side, MWF-processed signals may present small coherence coefficients (as will be shown later). In this situation, $IPD_{out}^v(k)$ estimates may present large variance. In the limiting case, in which there is no coherence between residual noise components in $z_L(\lambda,k)$ and $z_R(\lambda,k)$ ($|\rho(k)| \to 0$), (14) turns to [34]

$$p_\theta(\theta) = \frac{1}{2\pi}. \quad (15)$$

In this situation, the phase has a uniform distribution. Therefore the performance of the estimator depends on the magnitude of the coherence coefficient $\rho$.

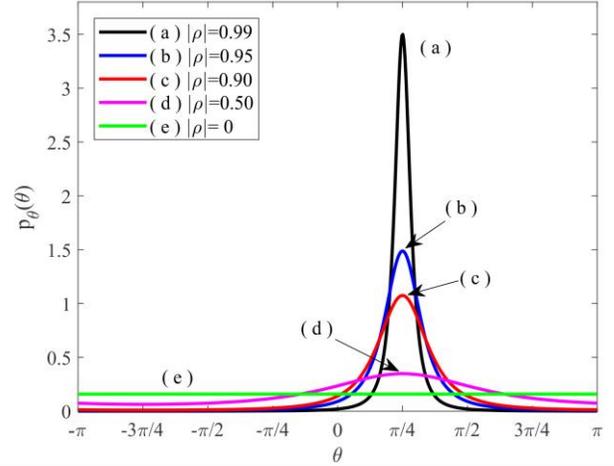

Fig. 2. Marginal probability density function for the phase of $v_L/v_R$ as a function of $|\rho| = |\mathbb{E}\{v_L v_R^*\}/(\sigma_{v_L}\sigma_{v_R})|$.

### B. Interaural Coherence

In consonance with results presented in Fig. 2, experiments presented in [35], applied to hearing aids, indicated strong relation between IC (specifically to $|\rho|$) and the capacity of listeners to discriminate small ITD changes. In other words, the authors conclude that IC should be considered cooperatively with the ITD to improve the localization of sounds in free field. Lately, in [27], the MWF-IC technique was proposed to provide noise reduction, but preserving the dispersive characteristic of diffuse sound fields. Its cost function was defined as:

$$J_C(k) = J_W(k) + \alpha_C(k) J_{IC}^v(k), \quad (16)$$

where

$$J_{IC}^v(k) = |IC_{out}^v(k) - IC_{in}^v(k)|^2, \quad (17)$$

in which the mean input and output noise IC are respectively defined as

$$IC_{in}^v(k) = \frac{\mathbf{q}_L^T \boldsymbol{\Phi}_{vv}(k) \mathbf{q}_R}{\sqrt{\mathbf{q}_L^T \boldsymbol{\Phi}_{vv}(k) \mathbf{q}_L \mathbf{q}_R^T \boldsymbol{\Phi}_{vv}(k) \mathbf{q}_R}}, \quad (18)$$

and

$$IC_{out}^v(k) = \frac{\mathbf{w}_L^H(k) \boldsymbol{\Phi}_{vv}(k) \mathbf{w}_R(k)}{\sqrt{\mathbf{w}_L^H(k) \boldsymbol{\Phi}_{vv}(k) \mathbf{w}_L(k) \mathbf{w}_R^H(k) \boldsymbol{\Phi}_{vv}(k) \mathbf{w}_R(k)}}. \quad (19)$$

Considering a single directional noise source in free field, the noise signal can be modeled as

$$\mathbf{v}(\lambda, k) = \mathbf{h}(k) v(\lambda, k) \quad (20)$$

where $v(\lambda,k)$ denotes the noise signal and $\mathbf{h} = [\mathbf{h}_L^T \ \mathbf{h}_R^T]^T = [h_{L,1}(\lambda,k) \ldots h_{L,M_L}(\lambda,k) \ h_{R,1}(\lambda,k) \ldots h_{L,M_R}(\lambda,k)]^T$ is the noise steering vector. It contains the acoustic transfer functions between the noise source and each of the $M$ microphones. In this way, the noise coherence matrix turns to:

$$\boldsymbol{\Phi}_{vv}(k) = \mathbb{E}\{\mathbf{v}(\lambda,k)\mathbf{v}^H(\lambda,k)\} = \sigma_v^2(k) \mathbf{h}(k)\mathbf{h}^H(k). \quad (21)$$

Using (21) in (18), after some manipulations, results in [28]

$$IC_{in}^v(k) = \frac{\mathbf{q}_L^T \mathbf{\Phi}_{vv}(k) \mathbf{q}_R}{\sqrt{\mathbf{q}_L^T \mathbf{\Phi}_{vv}(k) \mathbf{q}_R \mathbf{q}_R^T \mathbf{\Phi}_{vv}(k) \mathbf{q}_L}}$$
$$= \frac{\mathbf{q}_L^T \mathbf{\Phi}_{vv}(k) \mathbf{q}_R}{|\mathbf{q}_L^T \mathbf{\Phi}_{vv}(k) \mathbf{q}_R|} \quad , \quad (22)$$
$$= e^{j\angle \mathbf{q}_L^T \mathbf{\Phi}_{vv}(k) \mathbf{q}_R + 2\pi p(k)}$$
$$= e^{jIPD_{in}^v(k)}$$

The Taylor series for the complex exponential is given by

$$e^{jx} = \sum_{k=0}^{\infty} (-1)^k \left[ \frac{x^{2k}}{2k!} + j\frac{x^{2k+1}}{(2k+1)!} \right] . \quad (23)$$

Using its first order approximation in (22) it comes to:

$$IC_{in}^v(k) \cong 1 + jIPD_{in}^v(k) . \quad (24)$$

Using (20) in (19) and its result and (24) in (17) leads to

$$J_{IC}^v(k) \cong |IPD_{out}^v(k) - IPD_{in}^v(k)|^2 = J_{IPD}^v(k) . \quad (25)$$

Eq. (25) shows that minimization of the IC cost function for directional signals correspond to minimization of the IPD cost function. Given such revelation, we propose to apply the MWF-IC to the directional noise source case with the aim of controlling the ITD binaural cues of the processed noise.

## VI. EXPERIMENTAL SETUP

The performance of the MWF-IC ($J_C$) was assessed and compared to the MWF-ITD ($J_T$), MWF ($J_W$) and unprocessed signals under objective measures and psychoacoustic experiments for the case of one directional noise source. Simulations were performed with head-related impulse responses (HRIRs) obtained from a multichannel binaural database [36]. In this database, a manikin with a shape of a human head and torso, wearing two behind-the-ear hearing aids with 3 microphones each ($M_L = M_R = 3$) was positioned inside an anechoic chamber. The acoustic sources have zero elevation, corresponding to the transverse plane of the dummy head. All acoustic scenarios presented here are comprised by one speech source and one noise source placed at distances of 80 cm and 3 m (far-field), respectively, from the manikin. The speech source was situated at zero azimuth $\theta_S = 0°$, while the noise source is simulated at four different azimuths $\theta_N = \{ -60°, -30°, 30°, 60° \}$ in the front of the manikin. The negative azimuths correspond to the left-hand side of the sagital plane of the manikin, while the positive azimuths correspond to the opposite side.

The speech signal is a male voice selected from [37], containing a sentence of 2.7 seconds. The speech was convolved with the 0° azimuth HRIR and manually labeled to emulate and ideal voice activity detector (VAD), avoiding misclassification. The performance impact due to estimation errors of a real VAD is not approached in this study.

The noise signal was obtained by low-pass filtering white noise, to limit its energy up to 1.5 kHz (ITD range according to the Duplex Theory) [38] [39]. This noise signal was convolved with four different HRIRs, creating four distinct acoustic scenarios: $S_0N_{-60}$, $S_0N_{-30}$, $S_0N_{30}$ and $S_0N_{60}$[3]. The signal to noise ratio (SNR) of the contaminated signal was defined as 0 dB in the ear closest to the noise source (called "worst ear"), for all scenarios and experiments. The SNR and PESQ of the noisy (raw) speech for the four studied acoustic scenarios are presented in Table I.

The sampling frequency was set to $f_s = 16$ kHz and the input signals were transformed to the frequency domain by an $N = 256$ bin Short-Time Fourier Transform, with an analysis window of 128 samples, zero padding, and 50% of overlap. The transformed signals in the STFT domain were reconstructed by the weighted overlap-and-add method [40].

The noisy speech, speech and noise signals were processed by the coefficients obtained by applying the Broyden-Fletcher-Goldfarb-Shanno quasi-Newton optimization method [41] [42] to $J_C$ and $J_T$ cost functions, as well as by the theoretical solution to $J_W$. The weighting factor defined for a given technique is kept fixed for all bins $\alpha(k) = \alpha$.

The experiments with volunteers were approved by the Ethics Committee in Human Research, under certificate 49741615.2.0000.0121 CEP-UFSC. All volunteers involved have read and signed the written informed consent form.

TABLE I
OBJECTIVE MEASURES FOR THE NOISY INPUT SIGNAL.

|  | $S_0N_{-30}$ | $S_0N_{-60}$ | $S_0N_{30}$ | $S_0N_{60}$ |
|---|---|---|---|---|
| $PESQ_L$ | 1.1 | 1.1 | 1.2 | 1.2 |
| $PESQ_R$ | 1.2 | 1.2 | 1.1 | 1.1 |
| $SNR_L$ [dB] | 0 | 0 | 4.6 | 4.5 |
| $SNR_R$ [dB] | 4.6 | 4.5 | 0 | 0 |

### A. Objective Measures

Five objective measures were calculated for evaluating the performance of both MWF-IC and MWF-ITD methods: The signal to noise ratio (SNR), which measures the noise reduction; the intelligibility weighted gain in signal to noise ratio ($\Delta$ISNR) [43], which estimates the intelligibility of the speech signal; wideband perceptual evaluation of speech quality (PESQ) [44], which measures the overall quality of the enhanced speech signal [45]; the interaural time difference error ($\Delta$ITD) [8], calculated up to 1.5 kHz, which measures the preservation of ITD; and the Mean Square Coherence Error ($\Delta$MSC) [27], which measures the coherence variation between input and output signals. Sub-indexes were added to refer to speech (S), noise (N), and left (L) and right (R) ears.

Objective results obtained for scenarios $S_0N_{-30}$ and $S_0N_{-60}$ were very similar to the same scenarios in the opposite side. For this reason, only results for $S_0N_{30}$ and $S_0N_{60}$ are shown.

### B. Psychoacoustic Experiments

Experiments with volunteers were performed to evaluate the psychoacoustic aspects of the noise signal processed by both MWF-ITD and MWF-IC. They were conducted using a head-phone (Sennheiser HD 202) connected into a laptop.

---

[3] $S_{\theta_S}N_{\theta_N}$ means the speech source (S) is placed at $\theta_S$ degrees of azimuth and the noise source (N) is at $\theta_N$ degrees of azimuth with respect to the head midline (right azimuths are considered positive and left negative).

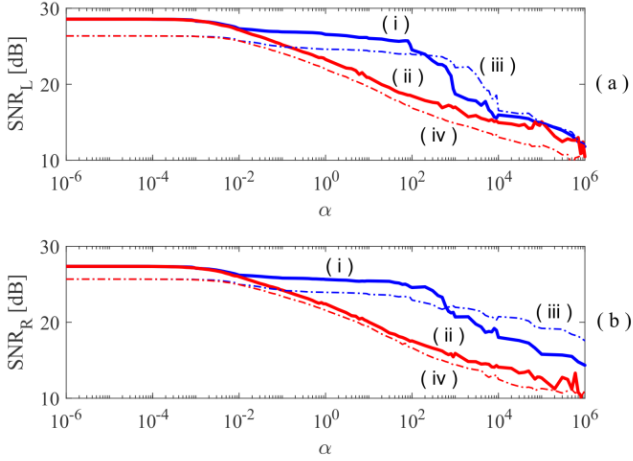

Fig. 3. SNR at the left (a) and right (b) ears for MWF-ITD (blue) and MWF-IC (red). $S_0N_{60}$ (thick continuous line) and $S_0N_{30}$ (thin dash-dotted line) scenarios: (i) $J_T$ for $S_0N_{60}$; (ii) $J_C$ for $S_0N_{60}$; (iii) $J_T$ for $S_0N_{30}$; (iv) $J_C$ for $S_0N_{30}$.

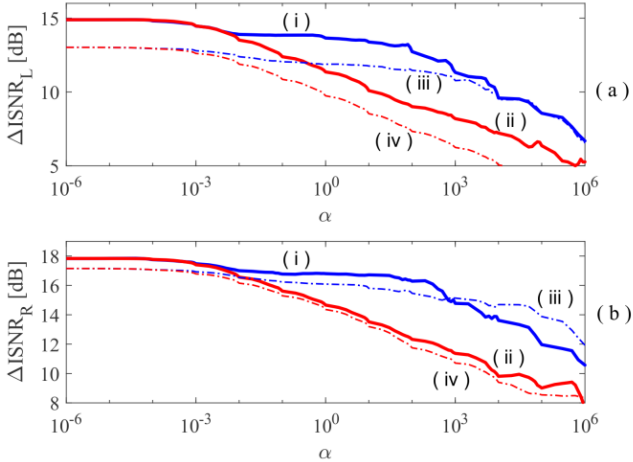

Fig. 4. $\Delta$ISNR at the left (a) and right (b) ears for MWF-ITD (blue) and MWF-IC (red). $S_0N_{60}$ (thick continuous line) and $S_0N_{30}$ (thin dash-dotted line) scenarios: (i) $J_T$ for $S_0N_{60}$; (ii) $J_C$ for $S_0N_{60}$; (iii) $J_T$ for $S_0N_{30}$; (iv) $J_C$ for $S_0N_{30}$.

The selected group of volunteers was comprised by 11 males and 4 females, with ages between 19 and 39 years-old (mean of 29 and standard deviation of 4.9 years-old). No previous complains regarding to hearing losses were declared. The experimental procedure was divided into 3 phases: (a) learning, (b) training, and (c) testing. In the first phase, the volunteers listened to the noise (only) signal filtered by HRIRs related to seven different azimuths { −90°, −60°, −30°, 0°, 30°, 60°, 90° }. Each audio was synchronized with visual information (presented on the screen of the laptop) related to the true azimuth of the processed noise signal. In the training phase, the volunteers were asked to identify the azimuth of the same 7 audios, presented in random order, without previous knowledge about the true azimuths. In this phase, volunteers which performed hemisphere inversions (lateralization errors related to the left-right sides) were drop out the experiment. In the test phase, the remaining volunteers were requested to classify a set of 16 audios, presented in random order. A virtual protractor with 13 combo boxes, ranged from −90° to 90° in steps of 15°, was presented in the screen of the laptop. The selected audios comprised four noise signals presented in the training phase (−60°, −30°, 30°, 60°), as well as filtered versions of the noise presented in the noisy speech, for the $S_0N_{-30}$, $S_0N_{30}$, $S_0N_{-60}$ and $S_0N_{60}$ scenarios, according to the optimum coefficients obtained from $J_W$, $J_T$ and $J_C$.

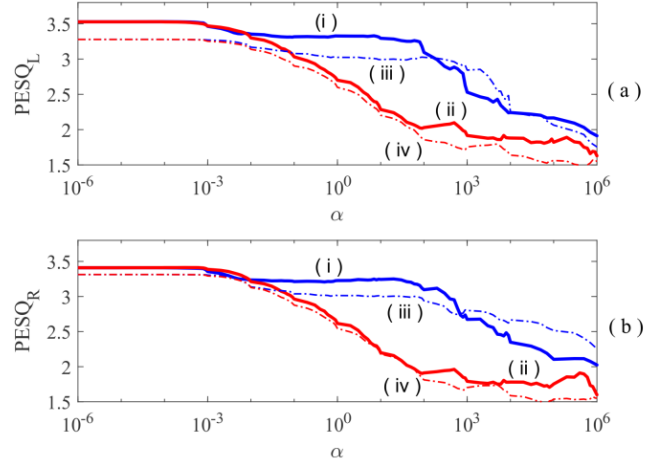

Fig. 5. PESQ at the left (a) and right (b) ears for MWF-ITD (blue) and MWF-IC (red), and $S_0N_{60}$ (thick continuous line) and $S_0N_{30}$ (thin dash-dotted line) scenarios: (i) $J_T$ for $S_0N_{60}$; (ii) $J_C$ for $S_0N_{60}$; (iii) $J_T$ for $S_0N_{30}$; (iv) $J_C$ for $S_0N_{30}$.

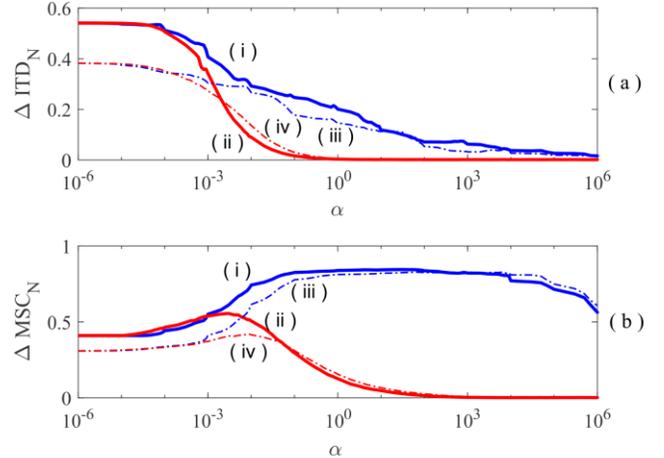

Fig. 6. Noise $\Delta$ITD (a) and noise $\Delta$MSC (b) for MWF-ITD (blue) and MWF-IC (red), and $S_0N_{60}$ (thick continuous line) and $S_0N_{30}$ (thin dash-dotted line) scenarios: (i) $J_T$ for $S_0N_{60}$; (ii) $J_C$ for $S_0N_{60}$; (iii) $J_T$ for $S_0N_{30}$; (iv) $J_C$ for $S_0N_{30}$.

TABLE II
WEIGHTING FACTORS FOR A MAXIMUM NOISE REDUCTION LOSS OF 15% AS COMPARED TO THE MWF TECHNIQUE

|  | $\alpha_T$ ($J_T$) | $\alpha_C$ ($J_C$) |
| --- | --- | --- |
| $S_0N_{-30}$ / $S_0N_{30}$ | $3\times10^3$ | 0.8 |
| $S_0N_{-60}$ / $S_0N_{60}$ | $4\times10^2$ | 0.4 |

For performance comparison purposes, a maximum noise reduction loss of 15%, as compared to the MWF solution, was deliberately set for both MWF-IC and MWF-ITD techniques in the "worst ear". This room establishes an arbitrary trade-off between noise reduction and spatial preservation. The weighting factors for attaining such specification, for both $J_C$ and $J_T$, are presented in Table II.

TABLE III
OBJECTIVE MEASURES FOR $S_0N_{30}$: $\alpha_T = 3\times10^3$ ($J_T$), $\alpha_C = 0.8$ ($J_C$)

|  |  | $J_W$ | $J_T$ | $J_C$ |
|---|---|---|---|---|
| $S_0N_{30}$ | $SNR_L$ [dB] | 26.3 | 21.3 | 22 |
|  | $SNR_R$ [dB] | 24.7 | 21.6 | 20.9 |
|  | $\Delta ISNR_L$ [dB] | 13.4 | 10.6 | 10.3 |
|  | $\Delta ISNR_R$ [dB] | 16.6 | 15.0 | 14.3 |
|  | $PESQ_L$ | 3.4 | 2.8 | 2.7 |
|  | $PESQ_R$ | 2.9 | 2.8 | 2.4 |
|  | $\Delta ITD_S$ | $2\times10^{-2}$ | $5\times10^{-2}$ | $2\times10^{-2}$ |
|  | $\Delta ITD_N$ | 0.4 | $3\times10^{-2}$ | $4\times10^{-3}$ |
|  | $\Delta MSC_S$ | $7\times10^{-3}$ | $5\times10^{-3}$ | $2\times10^{-2}$ |
|  | $\Delta MSC_N$ | 0.4 | 0.82 | 0.1 |

TABLE IV
OBJECTIVE MEASURES FOR $S_0N_{60}$: $\alpha_T = 400$ ($J_T$), $\alpha_C = 0.4$ ($J_C$)

|  |  | $J_W$ | $J_T$ | $J_C$ |
|---|---|---|---|---|
| $S_0N_{60}$ | $SNR_L$ [dB] | 28.5 | 22.7 | 24.2 |
|  | $SNR_R$ [dB] | 26.3 | 23.5 | 22.9 |
|  | $\Delta ISNR_L$ [dB] | 14.7 | 12.1 | 12.0 |
|  | $\Delta ISNR_R$ [dB] | 17.8 | 15.7 | 15.6 |
|  | $PESQ_L$ | 3.6 | 2.9 | 3 |
|  | $PESQ_R$ | 3.1 | 3 | 2.7 |
|  | $\Delta ITD_S$ | $2\times10^{-2}$ | $5\times10^{-2}$ | $3\times10^{-2}$ |
|  | $\Delta ITD_N$ | 0.43 | $7\times10^{-2}$ | $1\times10^{-2}$ |
|  | $\Delta MSC_S$ | $1\times10^{-2}$ | $1\times10^{-2}$ | $2\times10^{-2}$ |
|  | $\Delta MSC_N$ | 0.43 | 0.82 | 0.2 |

## VII. RESULTS

In this section, objective measures and psychoacoustic results are presented to assess the ITD preservation performance for both MWF-IC and MWF-ITD methods assuming the case of one directional speech source and one directional noise source.

### A. Objective Measures

Fig. 3 presents left and right SNR, as a function of the weighting factor ($\alpha$), for both MWF-IC and MWF-ITD for the $S_0N_{30}$ and $S_0N_{60}$ scenarios. Clearly, for both techniques, the SNR decreases with the increase of the weighting factor. The MWF-IC curves starts decreasing for a smaller $\alpha$ as compared to the MWF-ITD, which could lead us to prematurely disqualify the former. The plateau in the extreme left side of Fig. 3 corresponds to the SNR provided by the MWF ($\alpha \to 0$) technique. The same behavior is observed for both the $\Delta ISNR$ and PESQ, respectively presented in Figs. 4 and 5.

Fig. 6a presents $\Delta ITD$, which indicates that, as compared to the conventional MWF, both MWF-IC and MWF-ITD are capable of decreasing the input-output variation of the interau-ral time difference for increased weighting factors.

Fig. 6b shows that increasing the weighting factor of the MWF-ITD does not consistently reduce the mean square coherence error. By the other side, MWF-IC significantly reduces it, restoring the original interaural coherence of the residual noise.

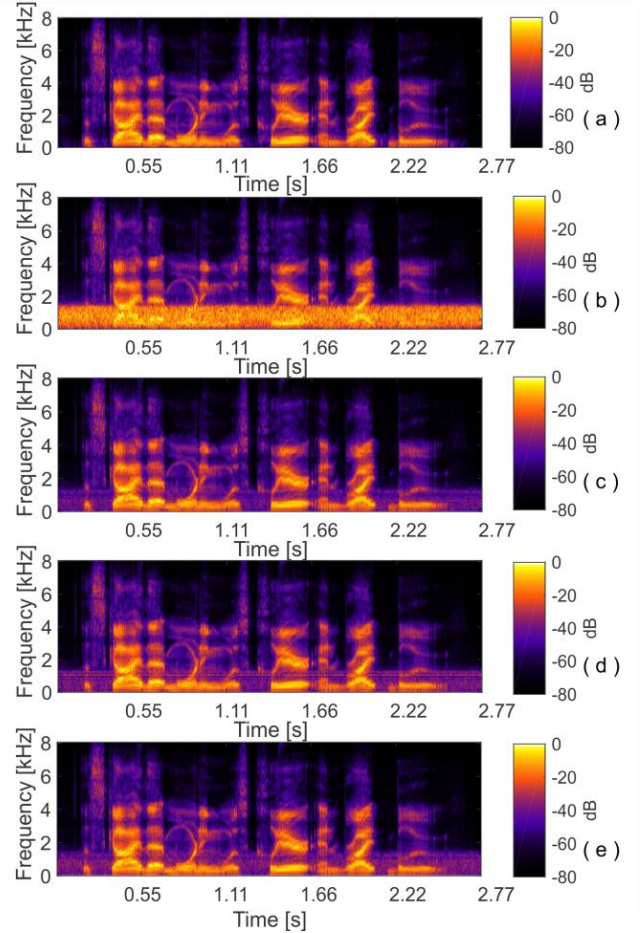

Fig. 7. Spectrograms of signals in the right ear for the $S_0N_{60}$ scenario: (a) clean speech; (b) contaminated speech; (c) MWF processed; (d) MWF-ITD processed; (e) MWF-IC processed.

### B. Psychoacoustic Experiments

Tables III and IV present the objective measures calculated for MWF, MWF-ITD and MWF-IC in the assessed scenarios ($S_0N_{30}$ and $S_0N_{60}$). In Fig. 7, the spectrograms of the analyzed signals, for the "best ear" and $S_0N_{60}$ scenario, are presented.

Fig. 8 shows results for the training phase of the 15 volunteers in the form of box-and-whisker diagrams. None of the volunteers performed lateralization inversions during the training phase. The box is represented by the blue rectangle that delimits the upper and lower quartiles denominated $q_1$ and $q_3$ (25% and 75% of values) respectively, while the red line inside the box means the median. In addition, the box represents the interquartile range (IQR), a measure of statistical dispersion between $q_3$ and $q_1$ (IQR = $q_3-q_1$). The whiskers show the lowest and highest sample values represented by the black dashed line. The outliers are reproduced with red crosses corresponding the samples greater than $q_3+\varpi(q_3-q_1)$ and lower than $q_3-\varpi(q_3-q_1)$. The variable $\varpi$ is defined as the default value of 1.5 [46] and represents the upper and lower extremes, which are not considered outliers.

volunteer ($\theta_S = 0°$) at $d_s$ meters. The noise source is indicated by a specific symbol with the corresponding true azimuth information. The box-and-whisker diagrams are represented by boxes filled with different colors according to the legends.

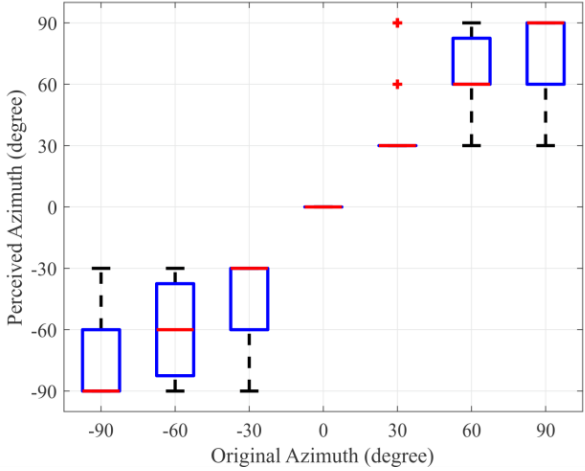

Fig. 8. Classification results obtained in the training phase for all 15 volunteers and low-pass noise.

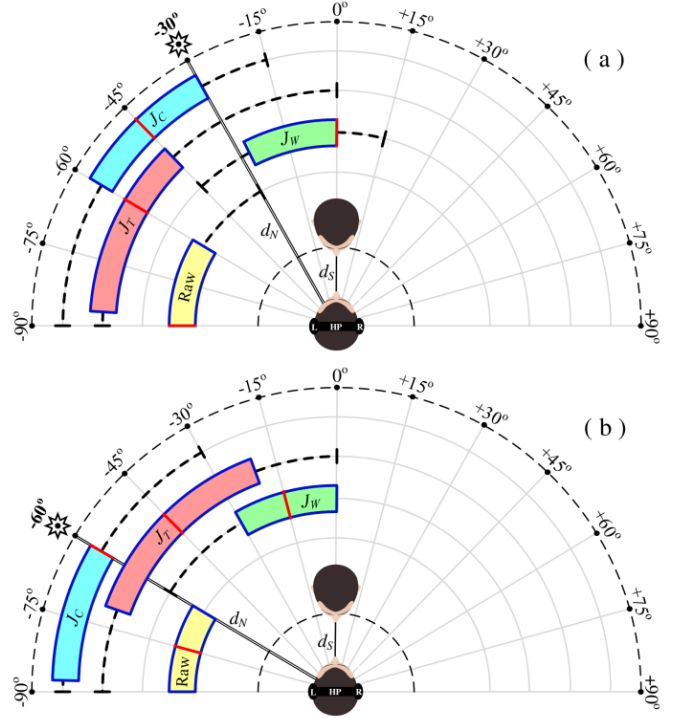

Fig. 10. Modified boxplot for the psychoacoustic experiment. Azimuth perception for the noise source. (a) $S_0N_{-30}$ (top), and (b) $S_0N_{-60}$ (bottom): Processed noise due to the MWF-ITD (red), MWF-IC (blue), MWF (green), and unprocessed noise (yellow).

TABLE V
PSYCHOACOUSTIC EXPERIMENT: AZIMUTH SAMPLE MEAN ($\bar{x}_i$), SAMPLE STANDARD DEVIATION ($s_i$) AND SAMPLE MEDIAN ($\tilde{x}_i$) FOR $i = \{R, W, T, C\}$, RESPECTIVELY MEANING: UNPROCESSED (RAW) AND PROCESSED BY MWF, MWF-ITD, AND MWF-IC TECHNIQUES.

|  |  | $S_0N_{-30}$ | $S_0N_{30}$ | $S_0N_{-60}$ | $S_0N_{60}$ |
|---|---|---|---|---|---|
| RAW | $\bar{x}_R$ | −74° | 65° | −75° | 62° |
|  | $s_R$ | 21° | 21° | 12.7° | 14.9° |
|  | $\tilde{x}_R$ | −90° | 75° | −75° | 60° |
| $J_W$ | $\bar{x}_W$ | −11° | 1° | −18° | −8° |
|  | $s_W$ | 19.2° | 9° | 19° | 27.7° |
|  | $\tilde{x}_W$ | 0° | 0° | −15° | 0° |
| $J_T$ | $\bar{x}_T$ | −59° | −27° | −44° | −53° |
|  | $s_T$ | 28.6° | 44° | 30° | 41.2° |
|  | $\tilde{x}_T$ | −60° | −15° | −45° | −75° |
| $J_C$ | $\bar{x}_C$ | −48° | 33° | −67° | 41° |
|  | $s_C$ | 22.1° | 27.1° | 19.5° | 22.3° |
|  | $\tilde{x}_C$ | −45° | 30° | −60° | 30° |

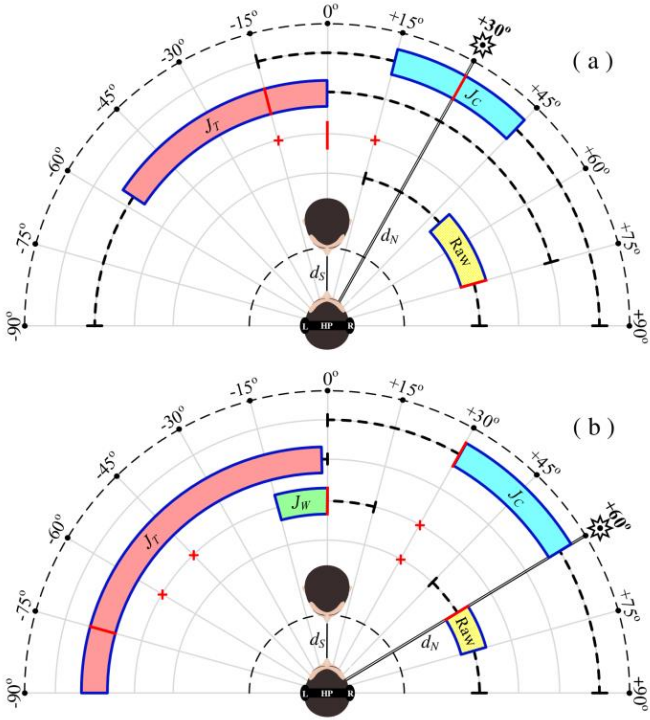

Fig. 9. Modified boxplot for the psychoacoustic experiment. Azimuth perception for the noise source. (a) $S_0N_{30}$ (top), and (b) $S_0N_{60}$ (bottom): Processed noise due to the MWF-ITD (red), MWF-IC (blue), MWF (green), and unprocessed noise (yellow).

Figs. 9 and 10 show the results for the 15 volunteers in the testing phase for the $S_0N_{-30}$, $S_0N_{30}$, $S_0N_{-60}$ and $S_0N_{60}$ scenarios. Modified box-and-whisker diagrams are presented for easiness of interpretation. This innovative design embodies the spatial perception into the box-and-whisker framework. Here, the volunteer location is represented by a sketch of a human head (top view) wearing a headphone. The speech source is represented by another sketch of human head in front of the

Table V shows the sample mean, sample standard deviation and sample median for boxplots presented in Figs. 9 and 10. Table VI shows the absolute differences between mean and median for the MWF-ITD and MWF-IC as compared to the original noise, for all scenarios.

TABLE VI
PSYCHOACOUSTIC EXPERIMENT: ABSOLUTE DIFFERENCES BETWEEN MEAN ($\bar{x}_i$) AND MEDIAN ($\tilde{x}_i$) AZIMUTHS FOR THE MWF-ITD (T) AND MWF-IC (C) AS COMPARED TO THE PERCEIVED AZIMUTH OF THE RAW (R) NOISE.

|  |  | $S_0N_{-30}$ | $S_0N_{30}$ | $S_0N_{-60}$ | $S_0N_{60}$ | Average |
|---|---|---|---|---|---|---|
| $J_T$ | $\|\bar{x}_R - \bar{x}_T\|$ | 15° | 92° | 31° | 115° | 63.3° |
|  | $\|\tilde{x}_R - \tilde{x}_T\|$ | 30° | 90° | 30° | 135° | 71.3° |
| $J_C$ | $\|\bar{x}_R - \bar{x}_C\|$ | 26° | 32° | 8° | 21° | 21.8° |
|  | $\|\tilde{x}_R - \tilde{x}_C\|$ | 45° | 45° | 15° | 30° | 33.8° |

The most impressive results were obtained for the MWF-ITD technique in both $S_0N_{30}$ and $S_0N_{60}$ scenarios. According to Fig. 9, nearly all volunteers performed lateralization inversions and, consequently, the sample means/medians pointed out to the opposite hemisphere of the true localization of the noise source. Aiming to clarify these results a new experiment was performed with the five best evaluators classified in the training phase. Firstly, they listened to 15 different realizations of the noise processed by the MWF-ITD and MWF-IC techniques for the $S_0N_{60}$ and $S_0N_{-60}$ scenarios. Following, they classified each signal according to the perceived hemisphere (L-left and R-right). Again, volunteers did not performed any lateralization inversions for the MWF-IC processed signals. By the other side, all volunteers presented hemisphere inversions for the MWF-ITD processed signals. Fig. 11 shows the results obtained for the MWF-ITD technique, in which the red dashed line indicates the average number of lateralization errors (6.4) for the 15 trials.

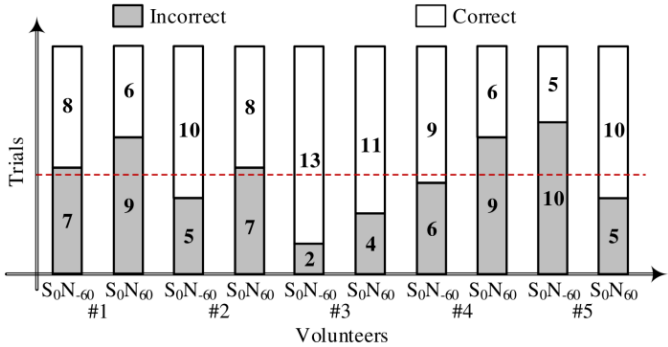

Fig. 11. Lateralization inversions obtained in the psychoacoustic experiments for the MWF-ITD technique: five volunteers and 15 trials.

## VIII. DISCUSSION

The objective measures SNR, ΔISNR, PESQ, and ΔITD presented in Figs. 3, 4, 5 and 6a clearly show a trade-off between noise reduction and spatial preservation as a function of the weighting factor. It means that acoustic comfort, intelligibility and quality are degraded, while the original noise ITD cues are preserved, with the increase of $\alpha$, for both MWF-ITD and MWF-IC methods. However, the ΔMSC, shown in Fig. 6b, indicates that MWF and MWF-ITD techniques may substantially change the interaural coherence of processed signals, independently of the choice of $\alpha$. Contrary, the MWF-IC consistently decreases both ΔITD and ΔMSC for increasing $\alpha$.

Tables III and IV present objective measures for both MWF-ITD and MWF-IC, considering weighting factors designed for a maximum noise reduction loss of 15% in the "worst ear" as compared to the MWF. The $SNR_R$ was decreased by 13% (MWF-ITD) and 15% (MWF-IC) for the $S_0N_{30}$ scenario ($\alpha_T = 3\times10^3$, $\alpha_C = 0.8$), and 11% (MWF-ITD) and 13% (MWF-IC) for $S_0N_{60}$ ($\alpha_T = 400$, $\alpha_C = 0.4$). The difference between the resulting $SNR_R$ for both methods did not exceed the just noticeable difference of 3 dB [47]. In the same way, both MWF-ITD and MWF-IC present approximately the same quality in the "worst ear" for both scenarios, since differences smaller than 0.2 PESQ are not clearly noticeable by volunteers [48]. The speech binaural cues were kept undistorted ($ΔITD_S$ and $ΔMSC_S < 0.1$), while $ΔITD_N$ was considerably reduced from 0.4 to less than 0.1. The most significant difference between MWF-ITD and MWF-IC is related to the $ΔMSC_N$. The $ΔMSC_N$ of the MWF processed noise, which originally was calculated as 0.4 was increased to 0.8 by the MWF-ITD, while the MWF-IC decreased it to 0.1 for the $S_0N_{30}$ scenario and to 0.2 for the $S_0N_{60}$. This indicates that the MWF-ITD acts only over the ITD cues, while the MWF-IC method controls both ITD and IC binaural cues.

The spectrogram in Fig. 7 provides a big picture of the "best ear". It shows that expressive noise reduction is achieved by MWF, MWF-ITD and MWF-IC, without significant visual differences.

Fig. 8 attested that all 15 volunteers present adequate lateralization judgment capacity. The calculated sample median accurately agreed with the true azimuths, and no hemisphere inversion were performed.

The main results of the psychoacoustic experiments are shown in Figs. 9 and 10. As expected, the MWF results (green boxes) were biased to the speech source azimuth (0°) [11]. It is also possible to verify that the median azimuth of the unprocessed noise (yellow box) is biased to a bigger azimuth magnitude as compared to its true value. This may be explained by the natural agreement between ILD and ITD binaural cues, which may amplify the lateralization subjective perception. This phenomenon does not occur with the processed noise, since the analyzed techniques do not act over the ILD.

Two main observations can be emphasized: Firstly, the MWF-IC provided more accurate (median) and precise (IQR) estimates of the true azimuths as compared to MWF-ITD (see Table V), for all acoustic scenarios. Secondly, while the MWF-IC always provides correct hemisphere localization, the MWF-ITD results a significant number of hemisphere inversions as shown in Fig. 11.

The importance of IC for effective ITD cue preservation was previously analyzed in [35]. This work assessed the lateralization capacity of volunteers in reverberant environments for different values of $|IC^v_{in}|$. Acoustic noise in the effective band of the ITD ($< 1.5$ kHz) was investigated. It was observed approximately 50% of hemisphere inversions for scenarios with $|IC^v_{in}| < 0.2$, whereas no inversions were reported for $|IC^v_{in}| > 0.8$. These results show the straight relation between the magnitude of the interaural coherence and the human abil-ity to lateralize sounds in free-field using the ITD information. An equivalent phenomenon can be observed in the MWF-IC results presented here. Fig. 12 shows an example of the magnitude of the interaural coherence as a function of the frequency for the noise signal applied in our experiments. The large coherence magnitude of unprocessed noise $|IC^v_{in}(k)|$ (yellow line - □) is characteristic of a free-field scenario (anechoic environment). The MWF-IC processed noise (blue line - *) preserves the original magnitude of the interaural coherence, basically resulting in $|IC^v_{out}(k)| > 0.8$. However, both MWF and MWF-ITD techniques provide reduced magnitude coherences. In fact, MWF-ITD may result very small values, achieving $|IC^v_{out}(k)| < 0.2$ for high frequencies. Such range of IC magnitudes is characteristic of diffuse acoustic fields, like those find in highly reverberant environments. The dashed dark lines inform the limit values for the magnitude of the interaural coherence described in [35].

In [49], psychoacoustic lateralization experiments indicated that small IC magnitudes increase the importance of ILD even in frequencies below 1.5 kHz. Frequencies containing dissonant ILD and ITD information may contribute to lateralization errors. The large number of hemisphere inversions verified with the MWF-ITD, may be attributed to the combination of small magnitude interaural coherence and discordant ILD and ITD binaural cues.

The obtained results indicate the MWF-IC is a promising technique for noise reduction with spatial preservation for both diffuse [50] and directional noise sources.

Finally, the simultaneous use of IC and ILD cost functions in (7) may lead to a noise reduction technique with perceptually relevant preservation of the spatial scenario along the entire hearing frequency range. This subject will be investigated in a future work.

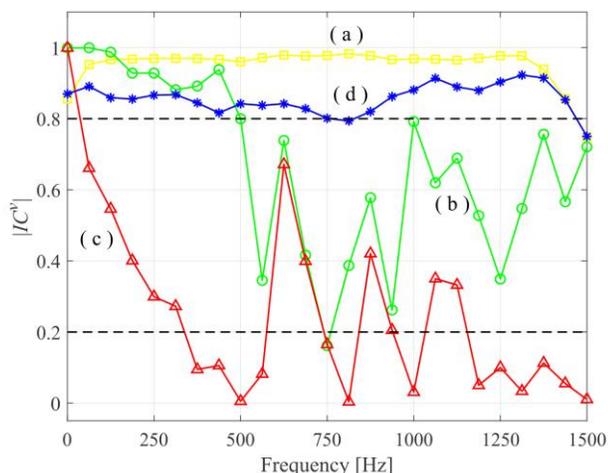

Fig. 12. Magnitude of the interaural coherence. (a) unprocessed noise (yellow square); (b) MWF processed noise (green circle); (c) MWF-ITD processed noise (red triangle); (d) MWF-IC processed noise (blue asterisk).

## IX. CONCLUSION

This paper proposes the use of the interaural coherence as a method for obtaining perceptually relevant preservation of interaural time differences in binaural hearing aids. It is shown that the MWF-ITD technique distorts the natural coherence presented by directional noise acquired by the microphones in free-field environments. The MWF-IC method provides a trade-off between noise reduction and preservation of the original noise spatialization for frequencies below 1.5 kHz. This happens due to the concomitant preservation of both ITD and IC binaural cues. Objective measures and psychoacoustic experiments corroborate the theoretical analysis, indicating that IC preservation is fundamental for ITD subjective perception and correct lateralization of directional sound sources in free-field. These results are of especially interest to hearing aid designers in search for a binaural noise reduction technique that preserves the original acoustic scenario for both diffuse and directional noise sources in frequencies below 1.5 kHz.

## APPENDIX

Considering the ratio of two correlated circularly-symmetric complex normal random variables

$$\psi = \psi_r + j\psi_i = x/y , \qquad (26)$$

with zero mean and complex correlation coefficient $\rho = \rho_r + j\rho_i = \mathbb{E}\{xy\}/(\sigma_x\sigma_y)$, $\sigma_x^2 = \mathbb{E}\{|x|^2\}$, $\sigma_y^2 = \mathbb{E}\{|y|^2\}$, their joint probability density function (PDF) in rectangular coordinates is defined as [34]

$$p_{\psi_r,\psi_i}(\psi_r,\psi_i) = \frac{1-|\rho|^2}{\pi\sigma_x^2\sigma_y^2}\left(\frac{|\psi|^2}{\sigma_x^2} + \frac{1}{\sigma_y^2} - 2\frac{\rho_r\psi_r + \rho_i\psi_i}{\sigma_x\sigma_y}\right)^{-2} . \quad (27)$$

Using polar coordinates

$$\psi_r = |\psi|\cos\theta , \quad \psi_i = |\psi|\sin\theta , \qquad (28)$$

in which $\theta$ denotes the phase of $\psi$. The joint PDF in polar coordinates is given by

$$p_{|\psi|,\theta}(|\psi|,\theta) = |\psi|\, p_{\psi_r,\psi_i}(\psi_r,\psi_i) . \qquad (29)$$

The marginal PDF of the phase is obtained by integrating (29) with respect to $|\psi|$

$$p_\theta(\theta) = \int_0^\infty p_{|\psi|,\theta}\, d|\psi| , \qquad (30)$$

resulting in

$$p_\theta(\theta) = \frac{1-|\rho|^2}{2\pi(1-\eta^2)}\left[\frac{\eta}{\sqrt{1-\eta^2}}\arccos(-\eta) + 1\right], \qquad (31)$$

in which

$$\eta = |\rho|\cos(\angle\rho - \theta) . \qquad (32)$$

### REFERENCES

[1] N. A. Whitmal, J.C. Rutledge, J. Cohen, "Reducing correlated noise in digital hearing aids," *IEEE Eng. Med. Biol. Mag.*, vol. 15, no. 5, pp. 88-96, 1996.
[2] V. Hamacher, "Comparison of advanced monaural and binaural noise reduction algorithms for hearing aids," in *Proc. IEEE Int. Conf. Acoust., Speech, Signal Process. (ICASSP)*, 2002, vol. IV, pp. 4008-4011.
[3] S. Kochkin, "MarkeTrak VIII: Consumer satisfaction with hearing aids is slowly increasing," *Hear. J.*, vol. 63, no. 1, pp. 19-32, 2010.
[4] T. Bogaert, T. J. Klasen, M. Moonen, L. Deun, J. Wouters, "Horizontal localization with bilateral hearing aids: without is better than with," *J. Acoust. Soc. Am.*, vol. 119, no. 1, pp. 515-526, 2006.
[5] S. Doclo, W. Kellermann, S. Makino, S. E. Nordholm, "Multichannel signal enhancement algorithms for assisted listening devices: Exploiting

spatial diversity using multiple microphones," *IEEE Signal Process. Mag.*, vol. 32, no. 2, pp. 18-30, 2015.

[6] T. Bogaert, S. Doclo, J. Wouters, M. Moonen, "The effect of multimicrophone noise reduction systems on sound source localization by users of binaural hearing aids," *J. Acoust. Soc. Am.*, vol. 124, no. 1, pp. 484-497, 2008.

[7] S. Doclo, R. Dong, T. J. Klasen, J. Wouters, S. Haykin, M. Moonen, "Extension of the multi-channel Wiener filter with localization cues for noise reduction in binaural hearing aids," in *Proc. Int. Workshop Acoust. Echo Noise, Control (IWAENC)*, 2005, pp. 221-224.

[8] T. Bogaert, J. Wouters, S. Doclo, M. Moonen, "Binaural cue preservation for hearing aids using an interaural transfer function multichannel Wiener filter," in *Proc. IEEE Int. Conf. Acoust., Speech, Signal Process. (ICASSP)*, 2007, vol. IV, pp. 565-568.

[9] T. J. Klasen, S. Doclo, T. Bogaert, M. Moonen, J. Wouters, "Binaural multichannel Wiener filtering for hearing-aids: preserving interaural time and level-differences," in *Proc. IEEE Int. Conf. Acoustics, Speech, Signal Process. (ICASSP)*, 2006, vol. V, pp. 145-148.

[10] S. Eyndhoven, T. Francart, A. Bertrand, "EEG-informed attended speaker extraction from recorded speech mixtures with application in neuro-steered hearing prostheses," *IEEE Trans. Biomed. Eng.*, vol. 64, no. 5, pp. 1045-1056, 2017.

[11] B. Cornelis, S. Doclo, T. Bogaert, M. Moonen, J. Wouters, "Theoretical analysis of binaural multimicrophone noise reduction techniques," *IEEE Trans. Audio, Speech, Lang. Process.*, vol. 18, no. 2, pp. 342-355, 2010.

[12] B. Cornelis, M. Moonen, J. Wouters, "Performance analysis of multichannel Wiener filter-based noise reduction in hearing aids under second order statistics estimation errors," *IEEE Trans. Audio, Speech, Lang. Process.*, vol. 19, no. 5, pp. 1368-1381, 2011.

[13] S. Doclo, M. Moonen, "On the output SNR of the speech-distortion weighted multichannel Wiener filter," *IEEE Signal Process. Let.*, vol. 12, no. 12, pp. 809-811, 2005.

[14] M. L. Hawley, R. Y. Litovsky, J. F. Culling, "The benefit of binaural hearing in a cocktail party: Effect of location and type of interferer," *J. Acoust. Soc. Am.*, vol. 15, no. 2, pp. 833-843, 2004.

[15] T. J. Klasen, T. Bogaert, M. Moonen, "Binaural noise reduction algorithms for hearing aids that preserve interaural time delay cues," *IEEE Trans. Signal Process.*, vol. 55, no. 4, pp. 1579-1585, 2007.

[16] S. Doclo, M. Moonen, "GSVD-based optimal filtering for single and multi-microphone speech enhancement," *IEEE Trans. Signal Process.*, vol. 50, no. 9, pp. 2230-2244, 2002.

[17] V. Benichoux, M. Rébillat, R. Brette, "On the variations of inter-aural time differences (ITDs) with frequency," in *Proc. Acoust.*, 2012, pp. 1-9.

[18] F. L. Wightman, D. J. Kistler, "The dominant role of low-frequency interaural time differences in sound localization," *J. Acoust. Soc. Am.*, vol. 91, no. 3, pp. 1648-1661, 1992.

[19] T. Francart, A. Lenssen, J. Wouters, "Enhancement of interaural level differences improves sound localization in bimodal hearing," *J. Acoust. Soc. Am.*, vol. 130, no. 5, pp. 2817-2826, 2011.

[20] V. Willert, J. Eggert, J. Adamy, R. Stahl, E. Körner, "A Probabilistic model for binaural sound localization," *IEEE Trans. Syst. Man Cybern. B: Cybern.*, vol. 36, no. 5, pp. 982-994, 2006.

[21] J. Middlebrooks, D. Green, "Sound localization by human listeners," *An. Rev. Psy.*, vol. 42, no. 1, pp. 135-159, 1991.

[22] W. Gaik, "Combined evaluation of interaural time and intensity differences: psychoacoustic results and computer modeling," *J. Acoust. Soc. Am.*, vol. 94, no. 1, pp. 98-110, 1993.

[23] M. H. Costa, P. A. Naylor, "ILD preservation in the multichannel Wiener filter for binaural hearing aid applications," in *Proc. Eur. Signal Process. Conf. (EUSIPCO)*, 2014, pp. 636-640.

[24] K. Blum, G.-J. Rooyen, H. A. Engelbrecht, "Spatial audio to assist speaker identification in telephony," in *Proc. Int. Conf. Syst. Signal Image Process. (IWSSIP)*, 2010, pp. 1-4.

[25] S. Doclo, R. Dong, T. Klasen, J. Wouters, S. Haykin, M. Moonen, "Extension of the multi-channel Wiener filter with ITD cues for noise reduction in binaural hearing aids," in *Proc Workshop Appl. Signal Proc Audio Acoust. (WAPSA)*, 2005, pp. 70-73.

[26] W. M. Hartmannn, E. J. Macaulay, "Anatomical limits in interaural time differences: an ecological perspective," *Front. Neurosci.*, vol. 8, no. 34, pp. 1-13, 2014.

[27] D. Marquardt, V. Hohmann, S. Doclo, "Coherence preservation in multi-channel Wiener filtering based noise reduction for binaural hearing aids," in *Proc. IEEE Int. Conf. Acoust., Speech, Signal Process. (ICASSP)*, 2013, pp. 8648-8652.

[28] E. Hadad, S. Doclo, S. Gannot, "The binaural LCMV beamformer and its performance analysis," *IEEE Trans. Audio, Speech, Lang. Process.*, vol. 24, no. 3, pp. 543-558, 2016.

[29] T. J. Klasen, M. Moonen, T. Bogaert, J. Wouters, "Preservation of interaural time delay for binaural hearing aids through multi-channel Wiener filtering based noise reduction," in *Proc. IEEE Int. Conf. Acoust., Speech, Signal Process. (ICASSP)*, 2005, pp. 29-32.

[30] D. Wang, G. Brown, *Computational Auditory Scene Analysis: Principles, Algorithms, and Applications*, Wiley-IEEE Press, pp. 155, 2006.

[31] S. Doclo, T. J. Klasen, T. Bogaert, J. Wouters, M. Moonen, "Theoretical analysis of binaural cue preservation using multi-channel Wiener filtering and interaural transfer functions," in *Proc. Int. Workshop Acoust. Echo Noise, Control (IWAENC)*, 2006, pp. 1-4.

[32] M. Raspaud, H. Viste, G. Evangelista, "Binaural source localization by joint estimation of ILD and ITD," *IEEE Trans. Audio, Speech, Lang. Process.*, vol. 18, no. 1, pp. 68-77, 2010.

[33] A. Papoulis, S. U. Pillai, *Probability, Random Variables and Stochastic Processes*. Fourth edition, McGraw-Hill, 2002.

[34] W.-J. Yan, W.-X. Ren, "Circularly-symmetric complex normal ratio distribution for scalar transmissibility functions. Part I: Fundamentals," *Mech. Syst. Signal Process.*, vol. 80, pp. 58-77, 2016.

[35] B. Rakerd, W. M. Hartmann, "Localization of sound in rooms. V. Binaural coherence and human sensitivity to interaural time differences in noise," *J. Acoust. Soc. Am.*, vol. 128, no. 5, pp. 3052-3063, 2010.

[36] H. Kayser, S. D. Ewert, J. Anemüller, V. Hohmann, B. Kollmeier, "Database of multichannel in-ear and behind-the-ear head-related and binaural room impulse responses," *EURASIP J. Adv. Signal Proc.*, no. 6, 2009.

[37] ITU-T, P.50, *Telephone Transmission Quality, Telephone Installations, Local Line Networks: Objective Measuring Apparatus – Artificial Voices*, Appendix I: test signals, 1998.

[38] T. T. Sandel, D. C. Teas, W. E. Feddersen, L. A. Jeffress,. "Localization of sound from single and paired sources," *J. Acoust. Soc. Am*., vol. 27, no. 5, pp. 842-852, 1955.

[39] A. W. Mills, "Lateralization of high-frequency tones," *J. Acoust. Soc. Am*., vol. 32, no. 1, pp. 132-134, 1960.

[40] R. E. Crochiere, "A weighted overlap-add method of short-time Fourier analysis/synthesis," *IEEE Trans. Acoust. Speech Signal Process.*, vol. 28, no. 1, pp. 99-102, 1980.

[41] J. S. Arora, *Introduction to Optimum Design*. Second edition. Elsevier, 2004.

[42] E. Habets, P. A. Naylor, "An online quasi-newton algorithm for blind SIMO identification," in *Proc. IEEE Int. Conf. Acoust., Speech, Signal* Process. (ICASSP), 2010, pp. 2662-2665.

[43] Y. Chisaki, K. Matsuo, K. Hagiwara, H. Nakashima, T. Usagawa, "Real-time processing using the frequency domain binaural model," *Appl. Acoust.*, vol. 68, pp. 923-938, 2007.

[44] J. V. Michalowicz, J. M. Nichols, F. Bucholtz, C. C. Olson, "A general Isserlis theorem for mixed-Gaussian random variables," *Stat. Prob. Lett.*, vol. 81, pp. 1233-1240, 2011.

[45] Y. Hu, P. C. Loizou, "Evaluation of objective quality measures for speech enhancement," *IEEE Trans. Audio Speech Lang. Process.*, vol. 16, pp. 229-238, 2008.

[46] D. Shannon, "Box-and-whisker plots with the SAS," *Pharm. Stat.*, vol. 2, pp. 291-295, 2003.

[47] D. McShefferty, W. M. Whitmer, M. A. Akeroyd, "The just-meaningful difference in speech-to-noise ratio," *Trends Hear.*, vol. 20, pp. 1-11, 2006.

[48] A. Servetti, J. C. Martin, "Error tolerant MAC extension for speech communications over 802.11 WLANs," in *Proc. Veh. Technol. Conf.* (VTC), 2005, pp. 1-5.

[49] W. M. Hartmann, B. Rakerd, Z. D. Crawford,. "Transaural experiments and a revised duplex theory for the localization of low-frequency tones," *J. Acoust. Soc. Am*., vol. 139, no. 2, pp. 968-985.

[50] D. Marquardt, S. Doclo, "Interaural coherence preservation for binaural noise reduction using partial noise Estimation and spectral postfiltering," *IEEE Trans. Audio, Speech, Lang. Process.*, vol. 26, no. 7, pp. 1261-1274, 2018.